\documentclass{PoS}

\usepackage{enumerate}
\usepackage{amsmath,amssymb}
\usepackage{mathrsfs}
\usepackage{bm}

\usepackage{ulem}
\usepackage{cleveref}

\title{
\begin{picture}(0,0)(0,0)%
 \put(360,105){\makebox(0,0)[l]{\textnormal{\normalsize
 CHIBA-EP-231}}}%
\end{picture}%
Mass-deformed Yang-Mills theory in the covariant gauge and its gauge-invariant extension through the gauge-independent BEH mechanism} 

\ShortTitle{Mass-deformed Yang-Mills theory}

\author{\speaker{Kei-Ichi Kondo}\\
Department of Physics, Graduate School of Science, Chiba University, Chiba 263-8522, Japan
\\
E-mail: \email{kondok@faculty.chiba-u.jp }}

\author{Yutaro Suda\\
Department of Physics, 
Graduate School of Science, 
Chiba University, Chiba 263-8522, Japan
}

\author{Masaki Ohuchi\\
Department of Physics, 
Graduate School of Science and Engineering, 
Chiba University, Chiba 263-8522, Japan
}

\author{Ryutaro Matsudo\\
Department of Physics, 
Graduate School of Science and Engineering, 
Chiba University, Chiba 263-8522, Japan
}

\author{Yui Hayashi\\
Department of Physics, 
Graduate School of Science and Engineering, 
Chiba University, Chiba 263-8522, Japan
}

\abstract{
In this talk we consider the mass-deformed Yang-Mills theory in the covariant gauge which is obtained by just adding a gluon mass term to the Yang-Mills theory with the covariant gauge fixing term and the associated Faddeev-Popov ghost term.
First, we reconfirm that the decoupling solution in the Landau gauge Yang-Mills theory is well reproduced from the mass-deformed Yang-Mills theory by taking into account loop corrections. 
Second, we show that the mass-deformed Yang-Mills theory is obtained as a gauge-fixed version of the gauge-invariantly extended theory which is identified with the gauge-scalar model with a single fixed-modulus scalar field in the fundamental representation of the gauge group.
This equivalence is obtained as a consequence of the gauge-independent Brout-Englert-Higgs mechanism proposed recently by one of the authors. 
Third, we show that the reflection positivity is violated for any value of the parameters in the mass-deformed Yang-Mills theory to one-loop quantum corrections. 
Finally, we discuss the implications for the existence of positivity violation/restoration crossover 
in light of the Fradkin-Shenker continuity between Confinement-like and Higgs-like regions in a single confinement phase in the gauge-scalar model on the lattice. 
}

\FullConference{XIII Quark Confinement and the Hadron Spectrum - Confinement2018\\
		31 July - 6 August 2018\\
		Maynooth University, Ireland}

\begin{document}

\section{Introduction}

It is still a challenging problem  in particle physics to explain quark and gluon confinement in the framework of quantum gauge field theories. 
%The very first question to this problem is to clarify what criterion should be adopted to understand confinement. 
The information on confinement is expected to be encoded in the gluon and ghost propagators which are obtained by fixing the gauge. 
Recent investigations have confirmed that in the Lorentz  covariant Landau gauge the \textbf{decoupling solution}  is the confining solution of the Yang-Mills theory in the three- and  four-dimensional spacetime, while the scaling solution is realized in the two-dimensional spacetime. 
Therefore, it is quite important to understand the decoupling solution in the Lorentz covariant Landau gauge. 
Of course, there are so many approaches towards this goal.  
In this talk, we focus on the approach \cite{TW10,Kondo15} which has been developed in recent several years and has succeeded to reproduce some features of the decoupling solution with good accuracy.  We call this approach the \textbf{mass-deformed Yang-Mills theory} with the gauge fixing term or the \textbf{massive Yang-Mills model} in the covariant gauge. 

However, the reason why this approach is so successful is not fully  understood yet in our opinion, although some attempts in this direction exist.  
In this talk  we show based on the previous paper \cite{Kondo16,Kondo18} that the mass-deformed Yang-Mills theory with the covariant gauge fixing term has the \textbf{gauge-invariant extension} which is given by a gauge-scalar model with a single  fixed-modulus scalar field in the fundamental representation of the gauge group, if a constraint which we call the \textbf{reduction condition} is satisfied. 
We call such a model the \textbf{complementary gauge-scalar model}. 
This equivalence is achieved based on the gauge-independent description \cite{Kondo16,Kondo18} of the \textbf{Brout-Englert-Higgs (BEH) mechanism} which does not rely on the spontaneous breaking of gauge symmetry.  
This description enables one to give a \textbf{gauge-invariant mass term of the gluon field} in the Yang-Mills theory which can be identified with the gauge-invariant kinetic term of the scalar field in the complementary gauge-scalar model.

In the lattice gauge theory, it is known that the confinement phase  in the pure Yang-Mills theory is analytically continued to the Higgs phase in the relevant gauge-scalar model, which is called the \textbf{Fradkin-Shenker continuity} \cite{FS79} as a special realization of the Osterwalder-Seiler theorem \cite{OS78}.
There are no local order parameters which can distinguish the confinement and Higgs phases. 
There is no thermodynamic phase transition between confinement and Higgs phases, in sharp contrast to the adjoint scalar case where there is a clear phase transition between the two phases. 
Therefore, Confinement and Higgs phases are just subregions of a single Confinement-Higgs phase \cite{FMS80,tHooft80}. 
%It should be remarked that these results do not deny the existence of non-local order parameters which can distinguish the confinement and the Higgs regions, see e.g. \cite{GM18}. 

It is known that the gluon propagator in the pure Yang-Mills theory exhibits the violation of \textbf{reflection positivity}. 
This fact was directly shown by the numerical simulations on the lattice, e.g., in the covariant Landau gauge \cite{CMT05,Bowmanetal07}.
We can show  analytically that the reflection positivity of the gluon \textbf{Schwinger function} is violated for any value of the parameters in the massive Yang-Mills  model with one-loop quantum corrections being included, see \cite{KSOMH18} for the details.
We discuss the implications for the existence of positivity violation/restoration crossover 
in light of the Fradkin-Shenker continuity between Confinement-like and Higgs-like regions in a single confinement phase in the gauge-scalar model on the lattice.

\section{Massive Yang-Mills theory and its gauge-invariant extension}
%\section{Massive Yang-Mills theory and its gauge-invariant extension}

%\begin{subequations}
We introduce the \textbf{mass-deformed Yang-Mills theory in the covariant gauge} which is defined just by adding the naive mass term $\mathscr{L}_{\rm m}$ to the ordinary massless Yang-Mills theory in the (manifestly Lorentz) covariant gauge $\mathscr{L}_{\rm YM}^{\rm tot}$,
%\begin{align}
%\mathscr{L}_{\rm mYM} 
%=  \mathscr{L}_{\rm YM}^{\rm tot} + \mathscr{L}_{\rm m}, \quad
%\mathscr{L}_{\rm m} = \frac12 M^2 \mathscr{A}^{\mu A}   \mathscr{A}_\mu^A ,
%= M^2 {\rm tr}(\mathscr{A}^\mu \mathscr{A}_\mu).
%\end{align}
%where 
%which we call the  \textbf{massive Yang-Mills model} in the covariant gauge   for short. 
%The ordinary massless Yang-Mills theory in the (manifestly Lorentz covariant)  Lorenz gauge with a gauge-fixing parameter $\alpha$ is given by
\begin{align}
\mathscr{L}_{\rm  YM}^{\rm tot} 
=&  \mathscr{L}_{\rm YM} + \mathscr{L}_{\rm GF} + \mathscr{L}_{\rm FP} + \mathscr{L}_{\rm m} , 
\nonumber\\
  \mathscr{L}_{\rm YM} =& -\frac{1}{4}\mathscr{F}^A_{\mu\nu}\mathscr{F}^{A\mu\nu} , 
\nonumber\\
\mathscr{L}_{GF}  =& \mathscr{N}^A \partial^\mu \mathscr{A}^A_\mu + \frac{\alpha}{2}\mathscr{N}^A\mathscr{N}^A ,
%\to -\frac{1}{2}\alpha^{-1} (\partial^\mu \mathscr{A}^A_\mu)^2 ,
%\quad (\alpha \to 0) , 
\nonumber\\
  \mathscr{L}_{\rm FP} =& i\mathscr{\bar C}^A \partial^\mu \mathscr{D}_\mu[\mathscr{A}]^{AB} \mathscr{C}^B 
%\nonumber\\
=  i\mathscr{\bar{C}}^A\partial^\mu(\partial_\mu\mathscr{C}^A+gf_{ABC}\mathscr{A}^B_\mu \mathscr{C}^C) ,
\nonumber\\
\mathscr{L}_m =& \frac{1}{2}M^2\mathscr{A}^A_\mu\mathscr{A}^{\mu A} ,
\end{align}
where $\mathscr{A}_\mu^A$ denotes the Yang-Mills field, $\mathscr{N}^A$ the Nakanishi-Lautrup field  and $\mathscr{C}^A, \mathscr{\bar{C}}^A$ ($A=1,...,{\rm dim}G$) the Faddeev-Popov ghost and antighost fields, which take their values in the Lie algebra $\mathscr{G}$ of a gauge group $G$ with the structure constants $f_{ABC}$ ($A,B,C=1,...,{\rm dim}G$). 
We call this theory the  \textbf{massive Yang-Mills model} in the covariant gauge for short.

It is shown \cite{KSOMH18} that the massive Yang-Mills model in a covariant gauge has the \textbf{gauge-invariant extension} given by the gauge-scalar model with a single radially-fixed (or fixed-modulus) scalar field in the fundamental representation of a gauge group if the theory is subject to an appropriate constraint which we call the \textbf{reduction condition}. 
We call such a gauge-scalar model the \textbf{complementary gauge-scalar model}. 
In other words, the complementary  gauge-scalar model with a single  {radially fixed fundamental scalar field}  reduces to the mass-deformed Yang-Mills theory in a fixed gauge if an appropriate reduction condition is imposed. 
For $G=SU(2)$, the complementary gauge-scalar model is given by
%with a single radially fixed fundamental scalar %subject to the reduction condition.
\begin{align}
\mathscr{L}_{\rm RF} =  \mathscr{L}_{\rm YM} + \mathscr{L}_{\rm kin} , \
%\nonumber\\ 
\mathscr{L}_{\rm YM} =  -\frac{1}{2} {\rm tr}[ \mathscr{F} _{\mu\nu} \mathscr{F}^{\mu\nu }  ], \
\mathscr{L}_{\rm kin} :=  ({D}_{\mu}[\mathscr{A}]\Phi )^{\dagger} \cdot ({D}^{\mu}[\mathscr{A}]\Phi ) , 
%\nonumber\\& 
%+ u(x) \left( \Phi ^{\dagger} \Phi  - \frac{1}{2}v^2   \right) ,
%\label{SU2-gauge-scalar-f1}
\end{align}
with a single fundamental scalar field subject to the radially fixed condition, 
%with the holonomic constraint,  
%$u(x)$ is the \textbf{Lagrange multiplier field} to incorporate the constraint:% that the radial degree of freedom or length of the scalar field is fixed $|\Phi(x)|=v/\sqrt{2}>0$: 
\begin{align}
		 f(\Phi(x)) :=  \Phi(x)^{\dagger} \cdot \Phi(x) - \frac{1}{2}v^2 = 0  ,
\label{SU2-s-constraint1}
\end{align}
where $\Phi (x)$ is the \textbf{$SU(2)$ doublet} 
%called the \textbf{Higgs doublet} which is 
formed from two complex scalar  fields   $\bm{\phi}_1 (x), \bm{\phi}_2 (x)$,
% which are parameterized (by the reason clarified later) as
\begin{align}
		\Phi(x) =&
	 \begin{pmatrix}
			\bm{\phi}_1(x) \\ 
			\bm{\phi}_2(x)
	\end{pmatrix}   , \
 \bm{\phi}_1(x), \bm{\phi}_2(x) \in \mathbb{C} , 
%\nonumber\\  
%=& \frac{1}{\sqrt{2}}
%		 \begin{pmatrix}
%			\phi_2(x) + i \phi_1(x) \\ 
%			\phi_0(x) - i \phi_3(x)
%	\end{pmatrix}  ,
%\nonumber\\  &
%  \ \phi_0(x), \phi_A(x) \in \mathbb{R} \ (A=1,2,3) .
\end{align} 
where $D_\mu[\mathscr{A}]$ is the covariant derivative in the fundamental representation $D_\mu[\mathscr{A}]:=\partial_\mu -ig  \mathscr{A}_\mu $. 
This theory is supposed to obey the  {reduction condition},
\begin{equation}
 \chi(x) := \mathscr{D}_\mu[\mathscr{A}]  \mathscr{W}^\mu(x) = 0 ,
\end{equation}
where $\mathscr{D}_\mu[\mathscr{A}]$ is the covariant derivative in the adjoint representation $\mathscr{D}_\mu[\mathscr{A}]:=\partial_\mu -ig [\mathscr{A}_\mu , \cdot ]$. 
Here $\mathscr{W}_\mu(x) = \mathscr{W}_\mu[\mathscr{A}(x), \Phi(x)]$ is the \textbf{massive vector field mode} defined shortly in terms of $\mathscr{A}_\mu$ and $\Phi$, which follows from the \textbf{gauge-independent BEH mechanism}. 
% with $M=gv/2$. 
%Here $\Phi (x)$ is the \textbf{$SU(2)$ doublet} 
%called the \textbf{Higgs doublet} which is formed from two complex scalar  fields   $\bm{\phi}_1 (x), \bm{\phi}_2 (x)$.
This gauge-scalar model is invariant under the gauge transformation, 
\begin{align}
 {\mathscr{A}}_\mu(x) &\to {\mathscr{A}}_\mu^{U}(x)  := U(x) \mathscr{A}_{\mu}(x) U(x)^\dagger + ig^{-1} U(x) \partial_\mu U(x)^\dagger , \
%\nonumber\\
\Phi(x)  \to \Phi^{U}(x)  := U(x) \Phi(x)  %, \ U(x) \in G 
 .
\end{align}
It is more convenient to convert the scalar field into the gauge group element. 
For this purpose, we introduce the \textbf{matrix-valued scalar field} $\Theta$ by adding another $SU(2)$ doublet $\tilde\Phi:=\epsilon \Phi^*$ as
\begin{align}
		\Theta(x) :=&
		  \begin{pmatrix} 
		\tilde\Phi(x) &	\Phi(x) 
	\end{pmatrix} 
=
		  \begin{pmatrix} 
		\epsilon \Phi^*(x) &	\Phi(x) 
	\end{pmatrix} 
%\nonumber\\
= 
		  \begin{pmatrix} 
		\bm{\phi}_2^*(x) &	\bm{\phi}_1(x) \\ 
		-\bm{\phi}_1^*(x) &	\bm{\phi}_2(x)
	\end{pmatrix} 
%\nonumber\\
%=&  \frac{1}{\sqrt{2}} (\phi_0 {\boldsymbol 1} + i \phi_A \sigma^A) 
%= \frac{1}{\sqrt{2}}
%	 \begin{pmatrix}
%			\phi_0 + i \phi_3 & \phi_2 + i \phi_1 \\ 
%			- \phi_2+i \phi_1  & \phi_0 - i \phi_3
%	\end{pmatrix}  
 , \
%\nonumber\\ 
  \epsilon =  
  \begin{pmatrix} 
		0 &	1 \\ 
		-1 & 0
	\end{pmatrix}
 .
	\label{Theta1}
\end{align} 
Then  we introduce the \textbf{normalized matrix-valued scalar field} $\hat{\Theta}$ by
\begin{align}
 \hat{\Theta}(x) =  {\Theta} (x)/\left(\frac{v}{\sqrt{2}}\right), \ v > 0   
.
\end{align} 
The above constraint (\ref{SU2-s-constraint1})  implies that the normalized scalar field $\hat{\Theta}$ obeys the conditions:
%$\hat{\Theta}(x) \hat{\Theta}(x)^\dagger=\bm{1}$.
$
\hat{\Theta}(x)^\dagger \hat{\Theta}(x) = \hat{\Theta}(x) \hat{\Theta}(x)^\dagger =  \bm{1} 
,
$
and
$
 \det \hat{\Theta}(x) 
= 1
$.
Therefore, the normalized matrix-valued scalar field $\hat{\Theta}$ is an element of $SU(2)$:
$
 \hat{\Theta}(x) \in G=SU(2) 
$.
This is an important property to give a gauge-independent BEH mechanism. 
The massive vector boson field $\mathscr{W}_\mu \in \mathscr{G}=su(2)$ is defined in terms of the original gauge field $\mathscr{A}_\mu \in \mathscr{G}=su(2)$ and  the normalized scalar field $\hat{\Theta} \in G=SU(2)$ as shown in a previous paper \cite{Kondo18},
\begin{align}
	\mathscr{W}_\mu(x) 
 :=& ig^{-1} ( {D}_{\mu}[\mathscr{A}] \hat{\Theta}(x)) \hat{\Theta}(x)^\dagger 
%\nonumber\\
=   -ig^{-1} \hat{\Theta}(x)	({D}_{\mu}[\mathscr{A}] \hat{\Theta}(x) )^\dagger
%\nonumber\\
%=& \frac{1}{2} ig^{-1} [ ( {D}_{\mu}[\mathscr{A}] \hat{\Theta}(x)) \hat{\Theta}(x)^\dagger -  \hat{\Theta}(x)	({D}_{\mu}[\mathscr{A}] \hat{\Theta}(x) )^\dagger ]
 .
\label{W1-SU2}
\end{align}
According to the gauge-independent BEH mechanism \cite{Kondo16,Kondo18}, the kinetic term of the scalar field $\Theta$ is identical to the mass term of $\mathscr{W}_\mu$,
%the mass term of $W$ is  
\begin{align}
 \mathscr{L}_{\rm kin}  =     \frac{1}{2}{\rm tr}(  ( {D}_{\mu}[\mathscr{A}] \Theta (x))^\dagger 	{D}^{\mu}[\mathscr{A}] \Theta(x) )  
%\nonumber\\
=  M^2{\rm tr}(\mathscr{W}_\mu(x) \mathscr{W}^\mu(x) ) , \ M=g\frac{v}{2} .
\label{kin=mass}
\end{align}

%\newpage
%%%%%%%%%%%%%%%%%%%%%%%%%%%%%%%%%%%%%%%%%%%%%%%%%%%%%%%%%%%%%
%%%%%%%%%%%%%%%%%%%%%%%%%%%%%%%%%%%%%%%%%%%%%%%%%%%%%%%%%%%%%
\section{
Massive Yang-Mills theory and decoupling solutions
}
%%%%%%%%%%%%%%%%%%%%%%%%%%%%%%%%%%%%%%%%%%%%%%%%%%%%%%%%%%%%%
%%%%%%%%%%%%%%%%%%%%%%%%%%%%%%%%%%%%%%%%%%%%%%%%%%%%%%%%%%%%%

\begin{figure}[t]
\centering
\includegraphics[width=6.5cm]{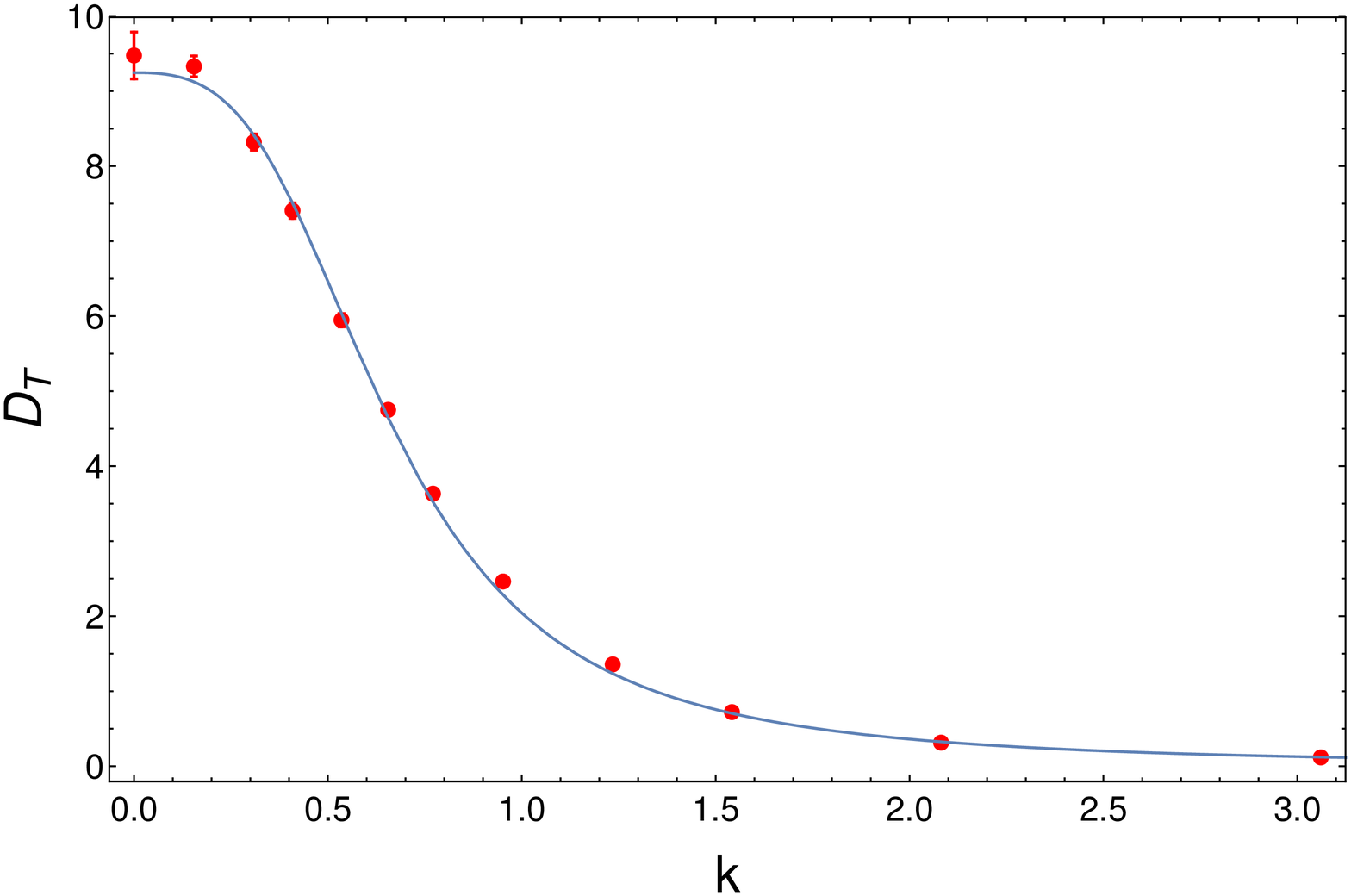}
\quad
\includegraphics[width=6.5cm]{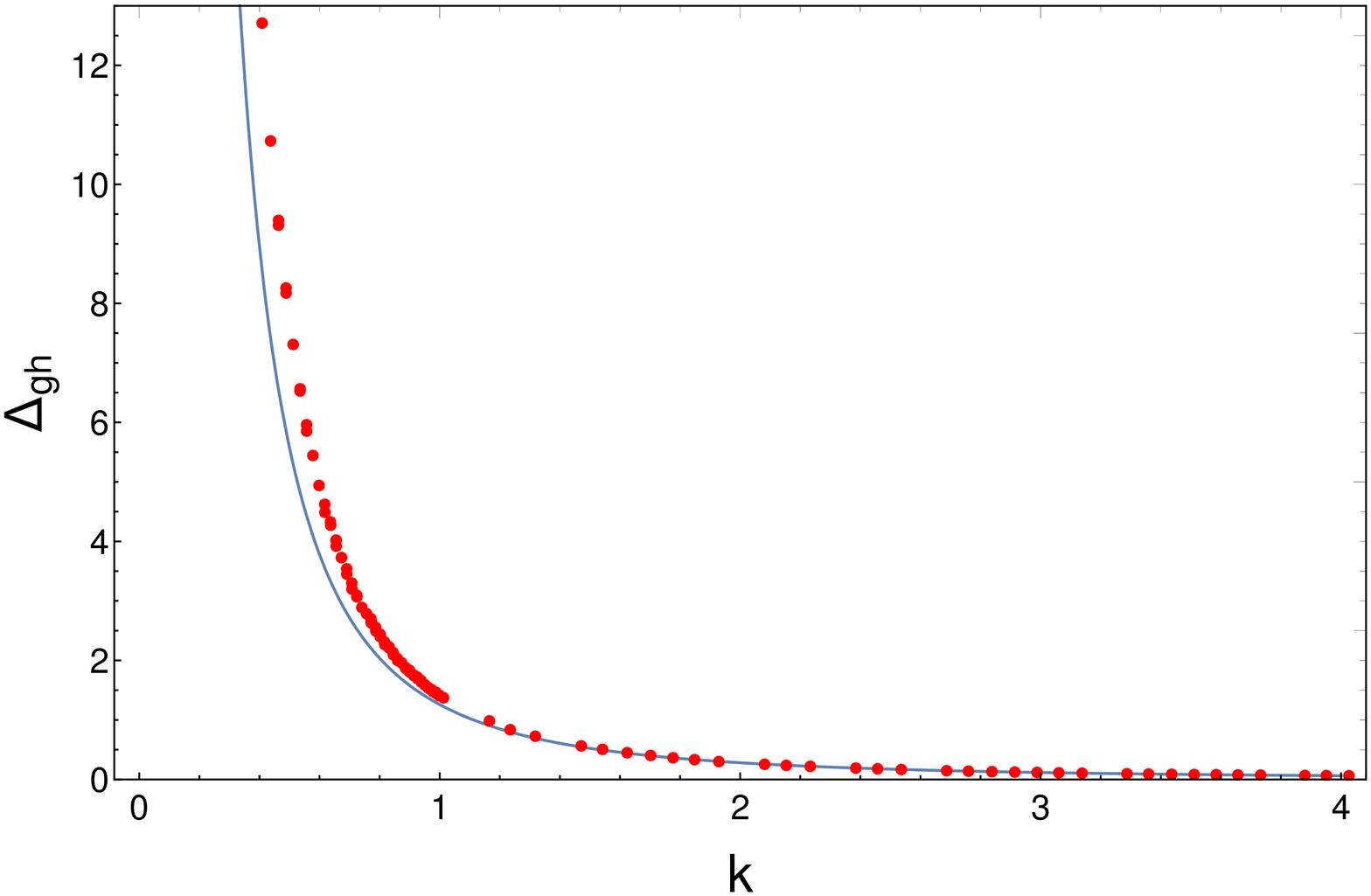}
\caption{\small
(Left) gluon propagator $\tilde D_T(k)$, (Right) ghost propagator $\Delta_{gh}(k)$ for the SU(3) Yang-Mills theory in the covariant Landau gauge. 
%The comparison of the analytical result with numerical simulations by Duarte,   Oliveira and Silva \cite{DOS16}
%[Duarte,   Oliveira and Silva (2016)]  
%on the lattice under the renormalization condition [OS] imposed at $\mu = 4$ GeV determines the fitting parameters as $g = 2.4$ and $M = 0.33$ GeV.
% \cite{bibOS}  
}
\label{fit}
\end{figure}

In order to reproduce the decoupling solution of the Yang-Mills theory in the covariant Landau gauge, we calculate one-loop quantum corrections to the gluon and ghost propagators in the massive Yang-Mills model.
The Nakanishi-Lautrup field $\mathscr{N}^A$ can be eliminated $\mathscr{L}_{GF} \to -\frac{1}{2}\alpha^{-1} (\partial^\mu \mathscr{A}^A_\mu)^2 
$. 
The results in the Landau gauge is obtained by taking the limit $\alpha \to 0$ in the final step of the calculations. 
Only in the Landau gauge $\alpha=0$ the massive Yang-Mills model with a naive mass term $\mathscr{L}_m$ has the gauge-invariant extension.
In order to obtain the gauge-independent results in the other gauges with $\alpha \not=0$, we need to include an infinite number of non-local terms in addition to the naive mass term $\mathscr{L}_m$ for gluons, as shown in \cite{KSOMH18}. %the previous section.   
Therefore, the results obtained in the Feynman gauge $\alpha=1$ from this theory must be different from those in the Landau gauge $\alpha=0$. 
We confirm that this is indeed the case by examining the positivity violation/restoration transition \cite{KSOMH18}. 
For comparison with the lattice data, we move to the Euclidean region and use $k_{E}$ to denote the Euclidean momentum.
We introduce the two-point vertex function $\Gamma_{\mathscr{A}}^{(2)}$ as the inverse of the propagator $\mathscr{D}_{\rm T}$ and the vacuum polarization function $\Pi_T$.  For gluons, we adopt the renormalization conditions.
 A renormalization condition adopted by  Tissier and Wschebor \cite{TW10} is  written as %in terms of $\Gamma$ or equivalently $\hat\Pi$ as%\cite{bibTW}:
\begin{align}
\label{TW}
%\text{[TW]} \quad
%\begin{cases}
%\Gamma_{\mathscr{A}}^{(2)}(k_{E} = 0) = M^2 
%\\
\Gamma_{\mathscr{A}}^{(2)}(k_{E} = \mu) = \mu^2 + M^2 ,
%\end{cases}
%  \Longleftrightarrow   
%\begin{cases}
%\hat{\Pi}^{\rm fin}(s = 0) = 0 
%\\
%\hat{\Pi}^{\rm fin}(s = \nu) = 0 \
%\end{cases}
%\nonumber\\  &
% ( \text{at} \ \mu = 1 \ \text{GeV} ). 
%\tag{\text{TW}}
\end{align} 
where $\mu$ is the renormalization scale.
%\begin{flalign}
%s:=\frac{k^2}{M^2}, \quad \nu:=\frac{\mu^2}{M^2} .
%\end{flalign}
For ghosts, we impose the renormalization condition: % which is common to both [TW] and [OS]:
\begin{align}
%\Delta_{gh}(k_{E} = \mu ) = \frac{1}{\mu^2} \Longleftrightarrow 
\Gamma_{gh}^{(2)}(k_{E} = \mu ) = \mu^2 .
%\nonumber\\  
%\Longleftrightarrow
%\hat{\Pi}_{gh}^{\rm fin}(s = \nu) = 0 .
\end{align}
The analytical results obtained by the massive Yang-Mills model (to one-loop order) was used to fit numerical simulations on the lattice  for the $SU(3)$ Yang-Mills theory in the Landau gauge. %\cite{DOS16}.
%[A.G. Duarte, O. Oliveira, and P.J. Silva, 
%Lattice Gluon and Ghost Propagators, and the Strong Coupling in Pure SU(3) Yang-Mills Theory: Finite Lattice Spacing and Volume Effects,  
%Phys. Rev. D{\bf 94} (2016)  014502.] 
%DOI: 10.1103/PhysRevD.94.014502 
See Fig.~\ref{fit}. 
%\newpage
\noindent
%\textcolor{green}{\Large\bf $\S$~ $SU(2)$ Yang-Mills from gauge-fundamental scalar  model }
%\setcounter{equation}{0}
%$\bigodot$ Fundamental scalar case: A perturbative approach
%\begin{figure}[!h]
%\centering
%\includegraphics[height=5.0cm]{fig-Higgs/gluon-propagator-Landau.eps}
%\hspace{5mm}
%\includegraphics[height=5.0cm]{fig-Higgs/ghost-propagator-Landau.eps}
%\centering\includegraphics[height=4.6cm]{SU3CGgauge12860.png}\hspace{5mm}\includegraphics[height=4.6cm]{SU3CGghost12860.png}
%\caption{
%\small
%(Left) gluon propagator $\tilde D_T(k)$, (Right) ghost propagator $\Delta(k)$.
%}
%\label{SU3CGgauge}
%\end{figure}
%arXiv:1605.00594 [hep-lat]] 
Both gluon and ghost propagators are well reproduced by a set of parameters $(g,M)$% for the renormalization condition [OS]:
%\begin{align}
%\text{[OS]:} \quad %g=2.47 , \quad M=0.329~\mathrm{GeV}.  
%& g  = 2.4 \pm 0.4 , 
%\quad 
%M  = 0.33 \pm 0.05 ~\mathrm{GeV} , 
%\notag \\
%& {\chi}^2_{reduced}  = 1.0009 ,
%\label{OS-parameter}
%\end{align}
%where ${\chi}^2_{\rm reduced}$ is defined as $\chi^2 / (d. o. f.)$, see Appendix~D.
%This result should be compared with that of Tissier and  Wschebor \cite{TW10} with the renormalization condition [TW]:
%[M. Tissier and N. Wschebor, 
%Infrared propagators of Yang-Mills theory from perturbation theory, 
%Phys.Rev. D{\bf 82}, 101701 (2010). 
%arXiv:1004.1607 [hep-ph]
%]
\begin{align}
%\text{[TW]:} \quad 
g=4.9, \quad M=0.54~\mathrm{GeV} .
\label{TW-parameter}
\end{align}
The more details for the choice of the renormalization condition are given in \cite{KSOMH18,HK18}. 

%\newpage
%%%%%%%%%%%%%%%%%%%%%%%%%%%%%%%%%%%%%%%%%%%%%%%%%%%%%%%%%%%%%
%%%%%%%%%%%%%%%%%%%%%%%%%%%%%%%%%%%%%%%%%%%%%%%%%%%%%%%%%%%%%
\section{
Positivity violation in the massive Yang-Mills model
}
%%%%%%%%%%%%%%%%%%%%%%%%%%%%%%%%%%%%%%%%%%%%%%%%%%%%%%%%%%%%%

%In this section, we show that the gluon propagator in the Yang-Mills theory leads to positivity violation suggesting gluon confinement.

%\subsection{Positivity violation  (Gluon confinement)}

%Reflection positivity
%In this section, we examine the reflection-positivity violation. 
The \textbf{Osterwalder-Schrader (OS) axioms}  are properties to be satisfied for the Euclidean quantum field theory formulated in the Euclidean space, which are the Euclidean version of the \textbf{Wightman axioms} for the relativistic quantum field theory formulated in the Minkowski spacetime. 
%\footnote{See the chapter of Millennium Prize Problems.
%\\ 
%K. Osterwalder and R. Schrader, 
%Commun. Math. Phys \textbf{31}, 83  (1973).
%Commun. Math. Phys \textbf{42}, 281 (1975). 
%\\
%The lattice version is given in: 
% K. Osterwalder  and E. Seiler,
% Annals Phys. \textbf{110}, 440 (1978).
%} 
A Wightman relativistic quantum field theory can be constructed from a set of Schwinger functions (Euclidean Green's functions) if they obey the Osterwalder-Schrader axioms.
In particular, the axiom of \textbf{reflection positivity} is the Euclidean counterpart to the positive definiteness of the norm in the Hilbert space of the corresponding Wightman quantum field theory.
If the positivity violation exists, a particular Euclidean correlation function  cannot have the  interpretation in terms of stable particle states, which is regarded as a manifestation of confinement. 
To demonstrate the violation of {reflection positivity} in the OS axioms, one counterexample suffices.
We consider a particular \textbf{Schwinger function} in the $D$-dimensional spacetime defined by the Fourier transform of the Euclidean propagator $\tilde{\mathscr{D}} (\bm{k}_{E} ,k_{E}^D)$
\begin{align}
\Delta(t) %:=& \Delta(\bm{k}_{E}=\bm{0},t)  
%:= \int d^{D-1}x   e^{-i\bm{k}_{E} \cdot \bm{x}}  \mathscr{D}( \bm{x},t)|_{\bm{k}_{E}=\bm{0}}
%\nonumber\\  
=& \int_{-\infty}^{+\infty} \frac{dk_{E}^D}{2\pi}  e^{ik_{E}^D t} \tilde{\mathscr{D}} (\bm{k}_{E}=\bm{0} ,k_{E}^D) .
\end{align} 
%If $\tilde{\mathscr{D}}(\bm{0}, k_{E}^D)$  is even in $k_{E}^D$, namely, $\tilde{\mathscr{D}}(\bm{0},-k_{E}^D) =\tilde{\mathscr{D}}(\bm{0},k_{E}^D)$,
%\footnote{
%\begin{align}
%  \Delta(t) 
%:=& \int d^dx \int \frac{d^Dp}{(2\pi)^D} e^{ipx} \tilde{\mathscr{D}}(p) 
%\nonumber\\
%=&  \int d^dx \int \frac{d^Dp}{(2\pi)^D} e^{i\bm{p} \cdot \bm{x}}  e^{ip_D t} \tilde{\mathscr{D}}(p)  
%\nonumber\\
%=&   \int \frac{d^Dp}{(2\pi)^D}  e^{ip_D t} \tilde{\mathscr{D}}(p)  \int d^dx e^{i\bm{p} \cdot \bm{x}} 
%\nonumber\\
%=&   \int \frac{d^Dp}{(2\pi)^D}  e^{ip_D t} \tilde{\mathscr{D}}(p)  (2\pi)^d \delta^d(\bm{p}) 
%\nonumber\\
%=&   \int_{-\infty}^{+\infty} \frac{dp_D}{2\pi}  e^{ip_D t} \tilde{\mathscr{D}}(\bm{p}=0,p_D)   
% ,
%\end{align}
%}
%the Schwinger function reduces  to 
%\begin{align}
%  \Delta(t) 
% = \int_{-\infty}^{+\infty} \frac{dk_{E}^D}{2\pi}  \cos (k_{E}^D t) \tilde{\mathscr{D}}(\bm{0} ,k_{E}^D)    .
%\end{align}
Therefore, non-positivity of the Schwinger function $\Delta(t)$   at some value of $t$ leads to  the  positivity violation. 
%\begin{align}
%\Delta(t) \le 0 \  \text{for some} \  t \ge 0
%\Longrightarrow
%\rho(\sigma^2) < 0 \  \text{for some} \ \sigma^2 \ge 0.
%\end{align}
Consequently, the reflection positivity is violated for the gluon propagator.
The corresponding states cannot appear in the physical particle spectrum, suggesting gluon confinement.

\begin{figure}[t]
\centering
\includegraphics[width=6cm]{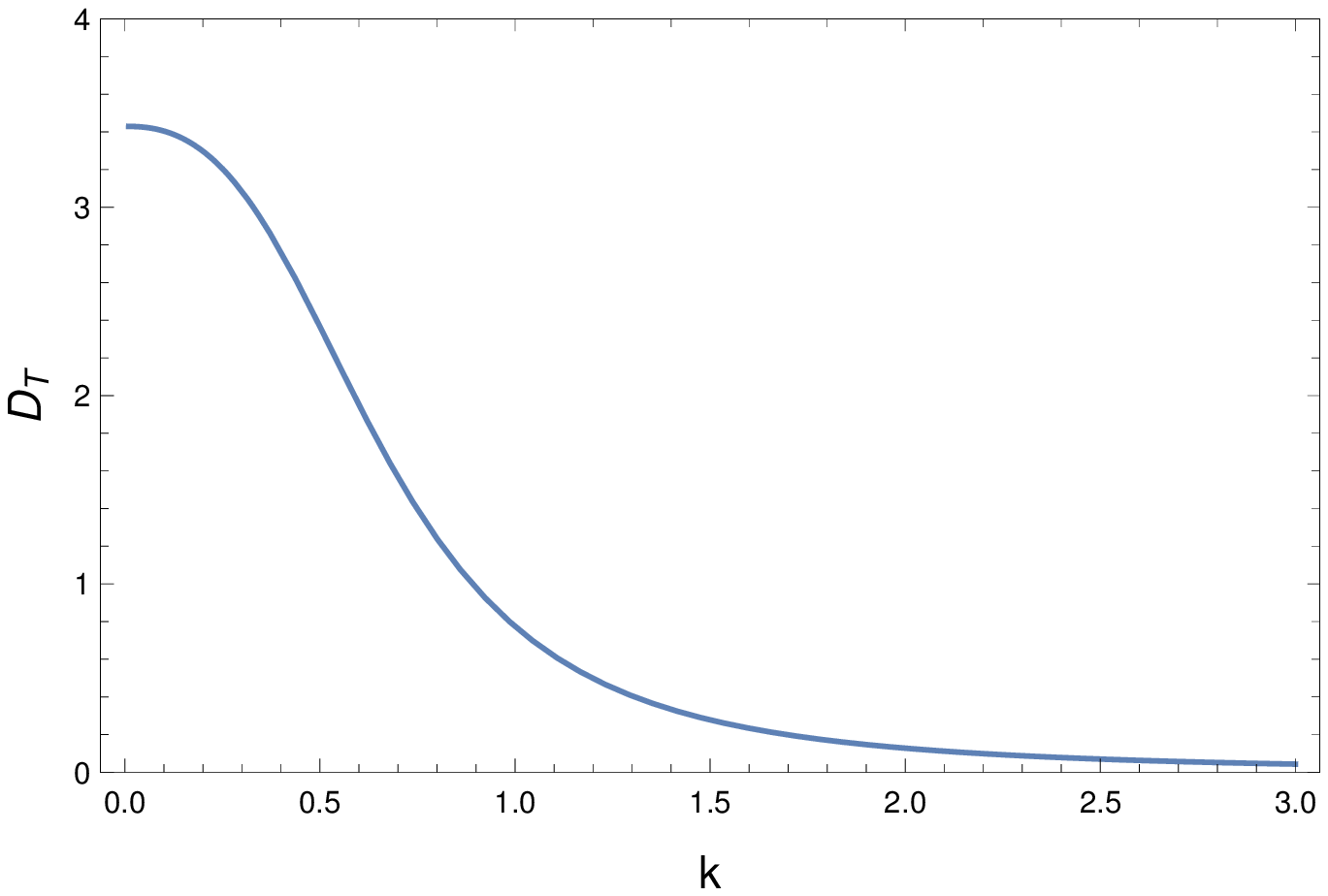}
\includegraphics[width=6cm]{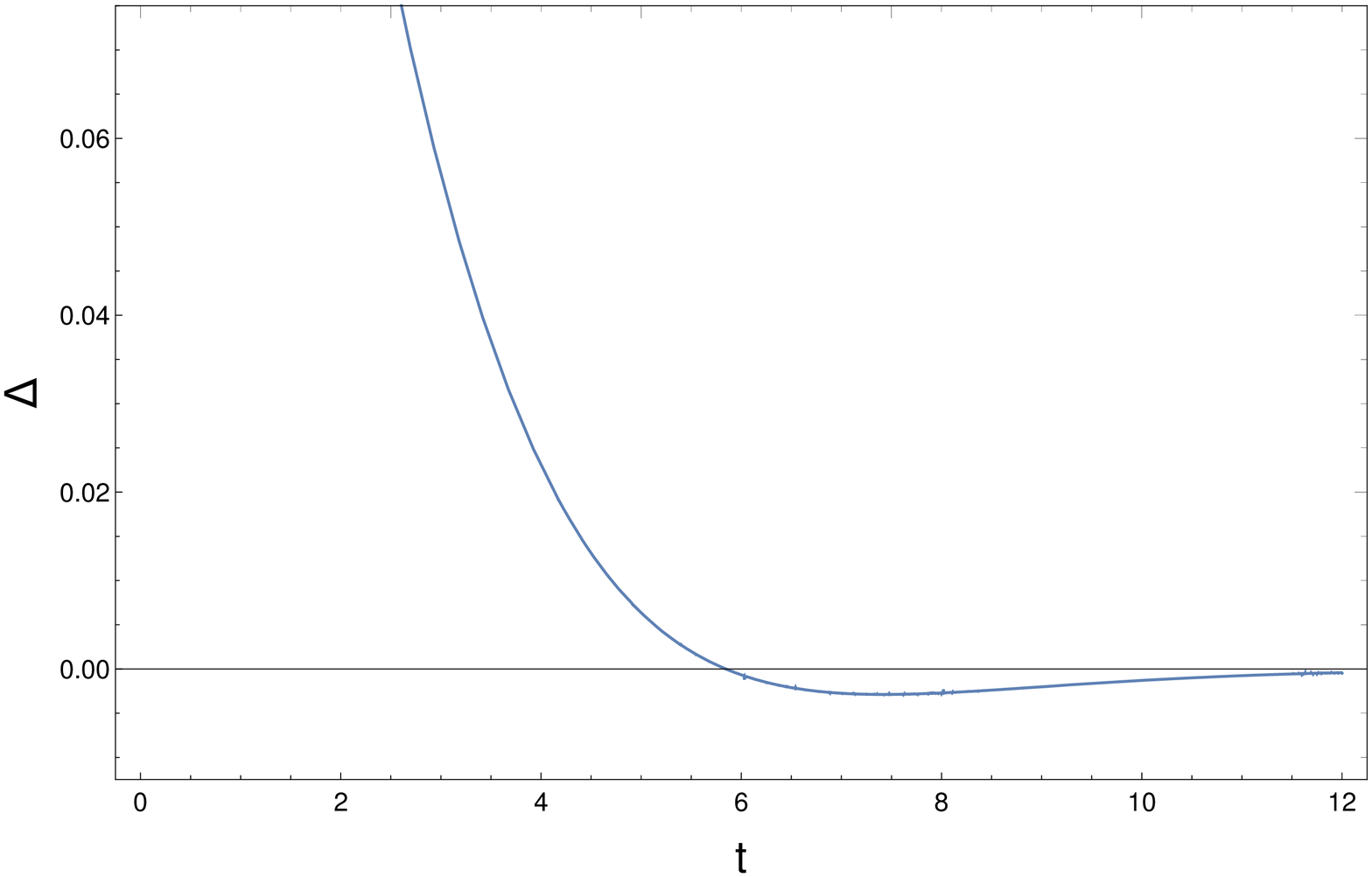}
\caption{\small
The gluon propagator $D_T$ and the resulting Schwinger function $\Delta$ in the covariant Landau gauge for the SU(3) Yang-Mills theory 
%under the renormalization condition [TW] 
with $g = 4.9$ and $M = 0.54$ GeV.
}
\label{lattice}
\end{figure}

Moreover, violation of reflection positivity  has been directly observed by measuring the Schwinger function on a lattice for $D=3$ \cite{CMT05}
%\footnote{A. Cucchieri, T. Mendes, and A. R. Taurines, 
%Phys. Rev. D\textbf{71}, 051902  (2005). 
%[hep-lat/0406020],
%}
 and for $D=4$, in both quenched (i.e., pure Yang-Mills theory) and unquenched (i.e., QCD) cases \cite{Bowmanetal07}. 
%\footnote{
%P.O. Bowman,  U.M. Heller,  D.B. Leinweber, M.B. Parappilly, A. Sternbeck, L. von Smekal, A.G. Williams and  Jian-bo Zhang,  
%Phys.Rev.D76, 094505 (2007). 
%[hep-lat/0703022], 
%}
In Fig.~\ref{lattice}, we give the plot for the gluon propagator and the resulting Schwinger function in the covariant Landau gauge  $\alpha=0$ for the SU(3) Yang-Mills theory 
%under the renormalization condition [TW] 
for $g = 4.9$ and $M = 0.54$ GeV.

%\begin{figure}[ht]
%\centering
%\includegraphics[width=9cm]{Ohuchi20180526/SchwingerCLTW-Lattice.eps}
%\includegraphics[width=9cm]{Ohuchi20180526/SchwingerCLTW-Lattice-Large.eps}
%\caption{\small
%The Schwinger function obtained from the gluon propagator in the covariant Landau gauge for the SU(3) Yang-Mills theory under the renormalization condition [TW] with the parameters $g = 4.9$ and $M = 0.54$ GeV.
%}
%\label{latticeLarge}
%\end{figure}

%\newpage
%%%%%%%%%%%%%%%%%%%%%%%%%%%%%%%%%%%%%%%%%%%%%%%%%%%%%%%%%%%%%
%%%%%%%%%%%%%%%%%%%%%%%%%%%%%%%%%%%%%%%%%%%%%%%%%%%%%%%%%%%%%
\section{
Positivity violation/restoration in the gauge-scalar model
}
%%%%%%%%%%%%%%%%%%%%%%%%%%%%%%%%%%%%%%%%%%%%%%%%%%%%%%%%%%%%%
%%%%%%%%%%%%%%%%%%%%%%%%%%%%%%%%%%%%%%%%%%%%%%%%%%%%%%%%%%%%%

\begin{figure}[t]
\centering
\includegraphics[width=6cm]{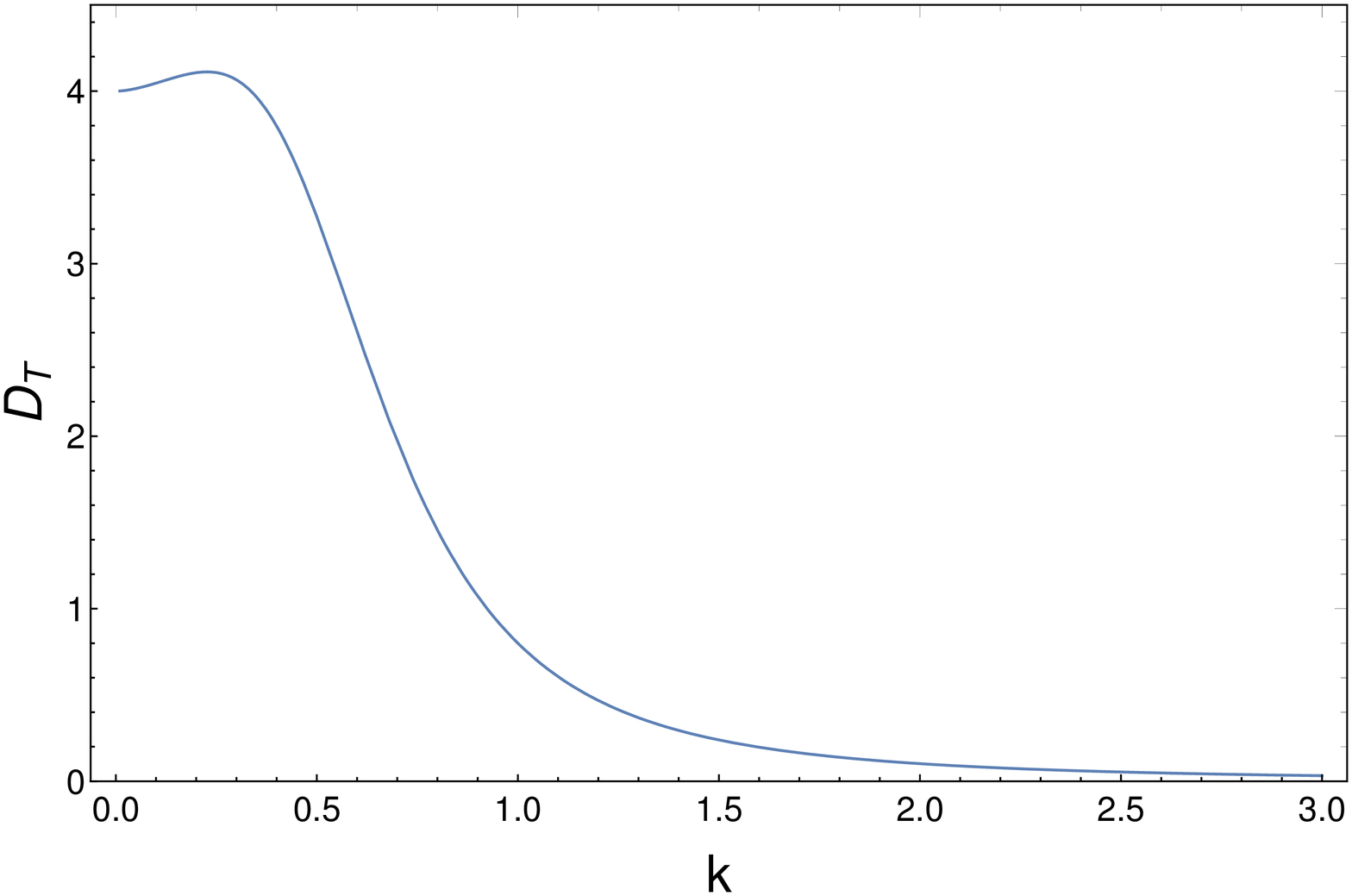}
\includegraphics[width=6cm]{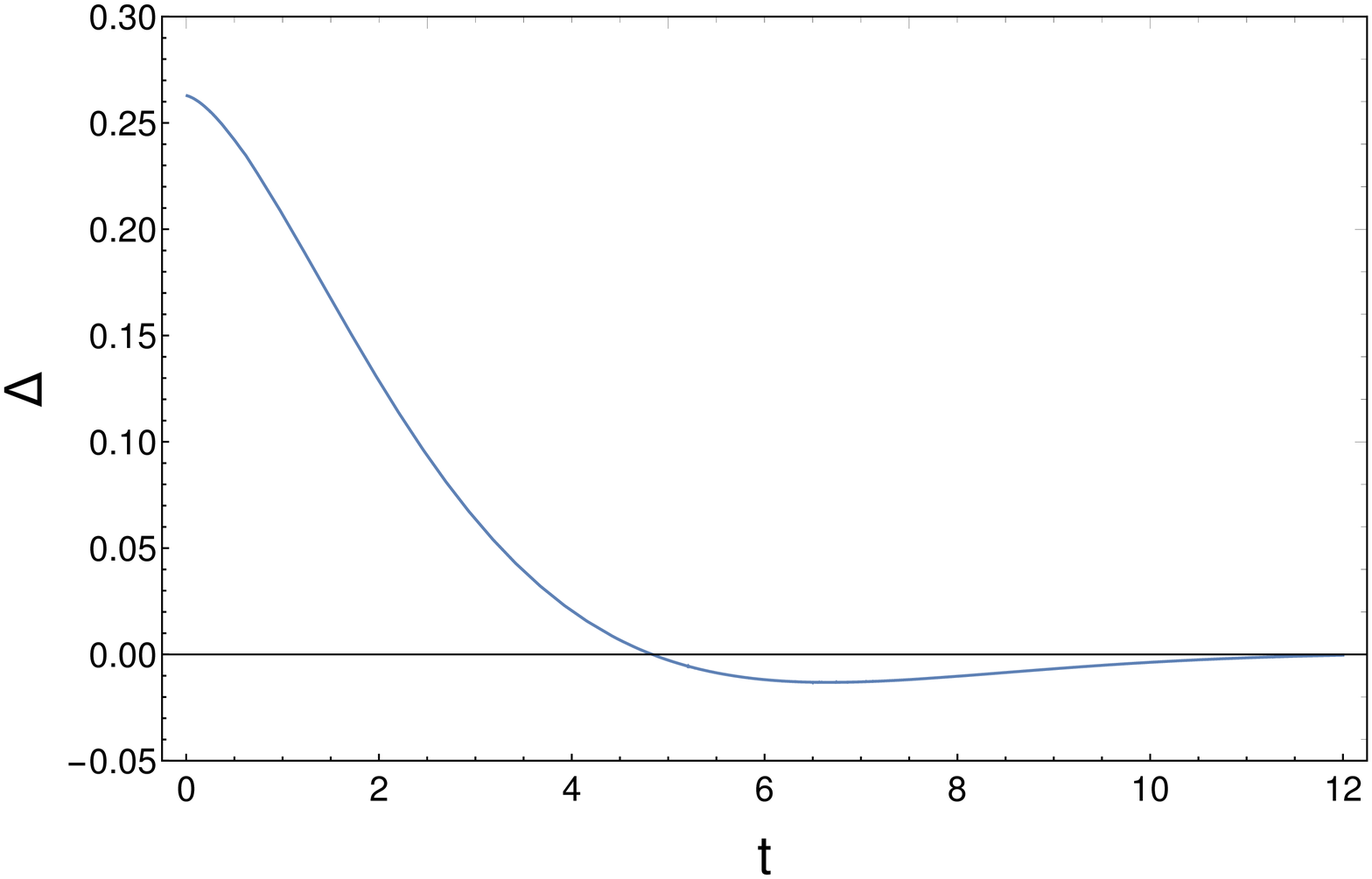}
\caption{\small
The gluon propagator $D_T$ and the resulting Schwinger function $\Delta$ in the covariant Landau gauge for the SU(3) Yang-Mills theory 
%under the renormalization condition [TW] 
with $g = 6$ and $M = 0.5$ GeV.
}
\label{nonPos}
\end{figure}

%\begin{figure}[t]
%\centering
%\includegraphics[width=6cm]{fig-ep231/MassivePropagatorCLTW-11-500.eps}
%\includegraphics[width=6cm]{fig-ep231/SchwingerCLTW-11-500.eps}
%\caption{\small
%The gluon propagator $D_T$ and the resulting Schwinger function $\Delta$ in the covariant Landau gauge for the SU(3) Yang-Mills theory under the renormalization condition [TW] with $g = 11$ and $M = 0.5$ GeV.
%}
%\label{singular}
%\end{figure}

\begin{figure}[t]
\centering
\includegraphics[width=6cm]{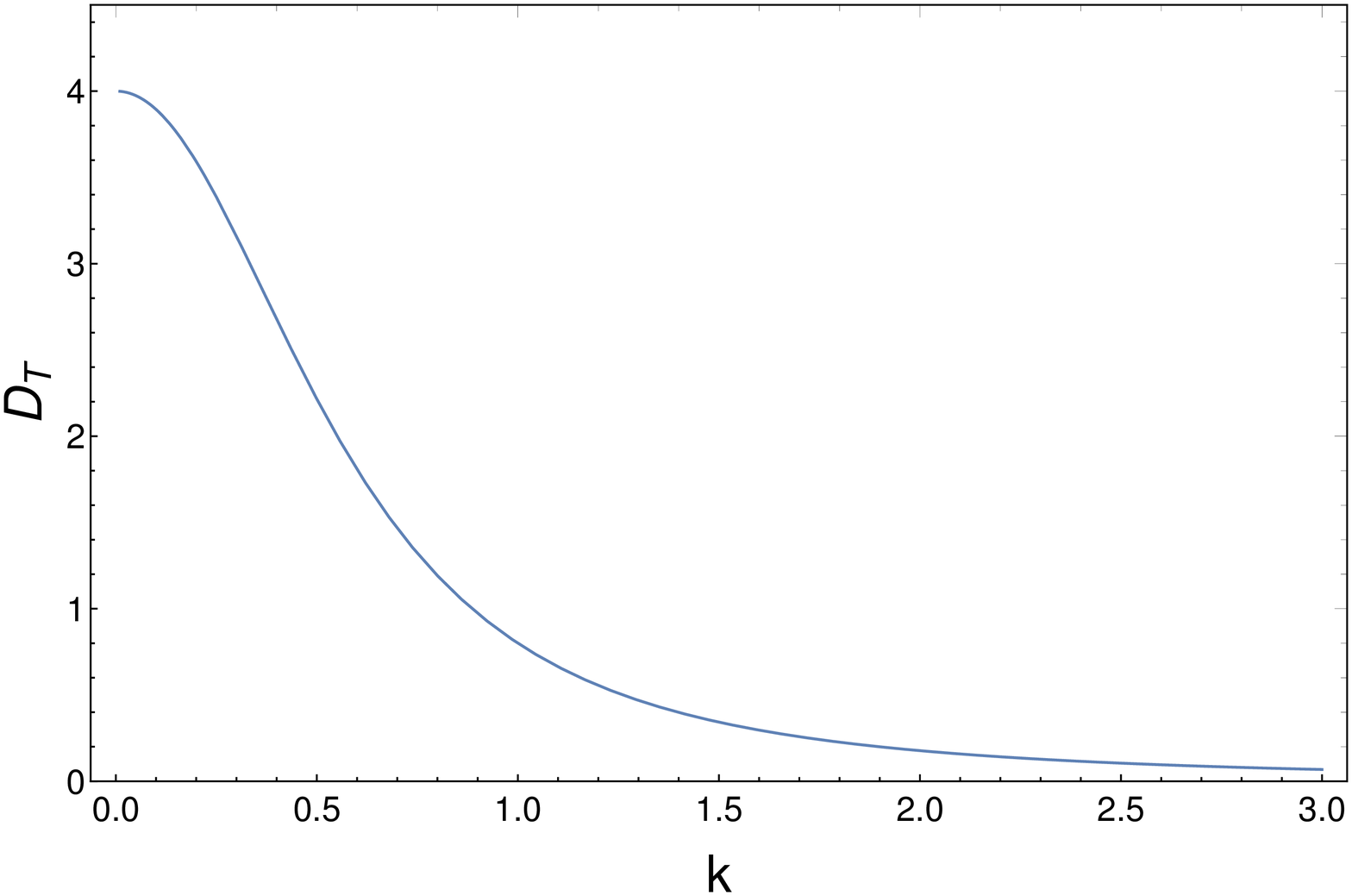}
\includegraphics[width=6cm]{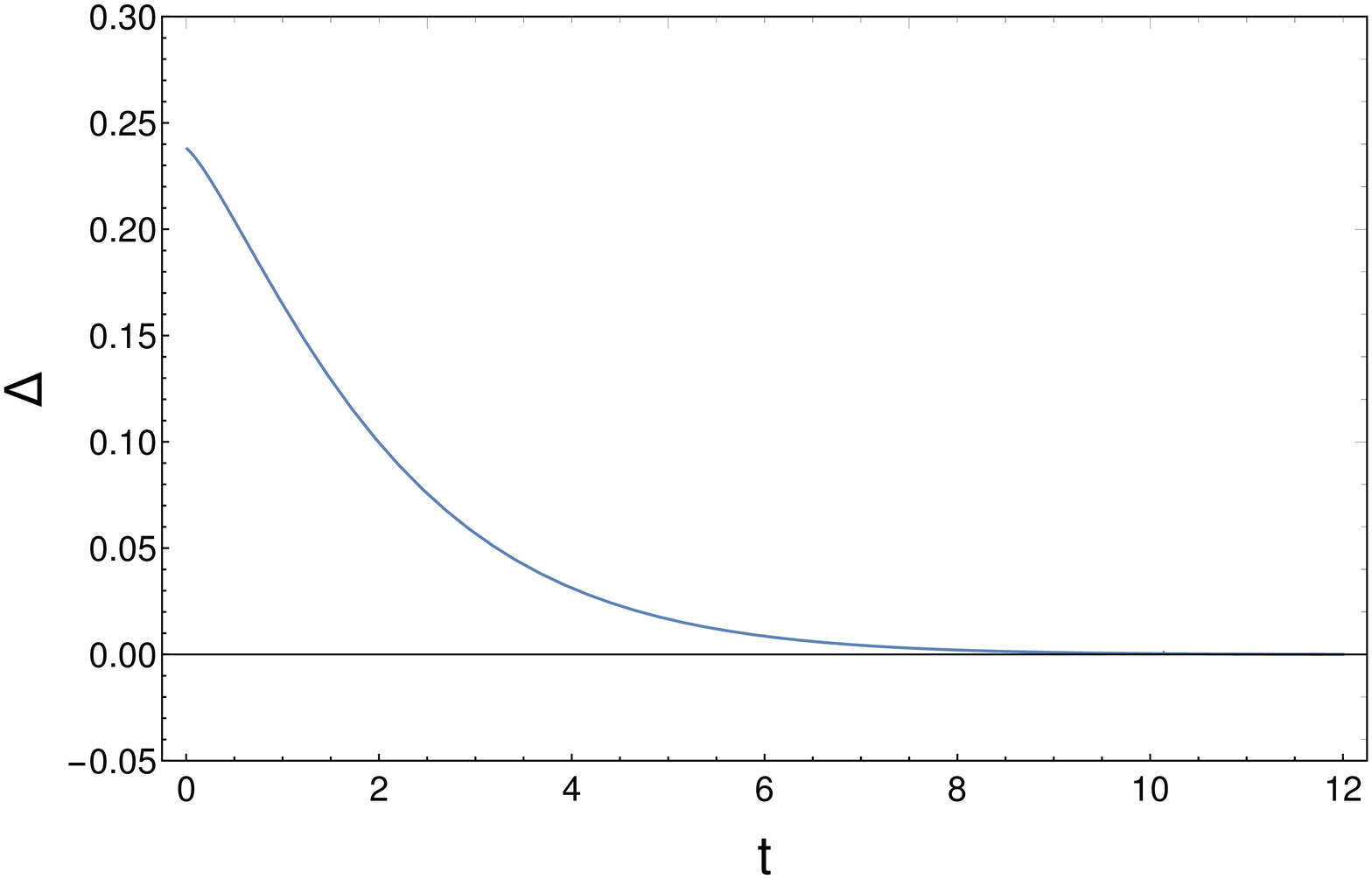}
\caption{\small
The gluon propagator $D_T$ and the resulting Schwinger function $\Delta$ in the covariant Landau gauge for the SU(3) Yang-Mills theory 
%under the renormalization condition [TW] 
with $g = 3$ and $M = 0.5$ GeV.
}
\label{pos}
\end{figure}

For larger coupling constant $g$ than the physical value  (\ref{TW-parameter}) in the Yang-Mills theory, the gluon propagator $\tilde{\mathscr{D}}_{\rm T}(p)$ exhibits non-monotonic behavior.
In fact, non-monotonicity of the propagator $\tilde{\mathscr{D}}(k^2)$ at some value of $k^2$ leads to  the  positivity violation. 
In Fig.~\ref{nonPos}, the gluon propagator is given for a different set of parameters $g = 6.0$ and $M = 0.5$ GeV. 
The gluon propagator clearly shows sizable non-monotonic behavior. 
Then the Schwinger function $\Delta(t)$ shows the stronger violation of positivity, which is to be compared with Fig.~\ref{lattice} for $g = 4.9$ and $M = 0.54$ GeV. 
For quite large coupling constant $g$, the gluon propagator $\tilde{\mathscr{D}}_{\rm T}(p)$ becomes singular at two values of $p$ and takes negative values in between.
%In Fig.~\ref{singular}, the gluon propagator is given for the parameters $g = 11$ and $M = 0.5$ GeV.
Therefore, this singular behavior affects the resulting Schwinger function  $\Delta(t)$. 
This feature will be an artifact due to the limitation of one-loop calculations. 
Therefore, we exclude this strong coupling region from the following considerations. 

For smaller coupling constant $g = 3$ and $M = 0.5$ than the physical value in the Yang-Mills theory, the gluon propagator $\tilde{\mathscr{D}}_{\rm T}(p)$ is monotonically decreasing in $p$ and the resulting Schwinger function seems to be positive as the numerical calculations demonstrate in Fig.~\ref{pos}. 
For smaller coupling constant, therefore, this numerical result seems to show that the positivity is not violated and is restored. 
In fact, if the positivity holds $\rho(\sigma^2) \ge 0$, the propagator $\tilde{\mathscr{D}}(k^2)$ is monotonically decreasing function of $k^2$. 
The result of Fig.~\ref{pos} seems to be consistent with this statement. 
However, it should be remarked that the monotonic propagator does not guarantee the positivity, since the converse is not necessarily true. %, see Appendix~E for the details. 
In fact, we can prove analytically that the reflection positivity of the gluon \textbf{Schwinger function} is violated for any value of the parameters $g $ and $M $ in the massive Yang-Mills  model with one-loop quantum corrections being included, see \cite{KSOMH18,HK18} for the details.

\begin{figure}[t]
\centering
\includegraphics[width=7cm]{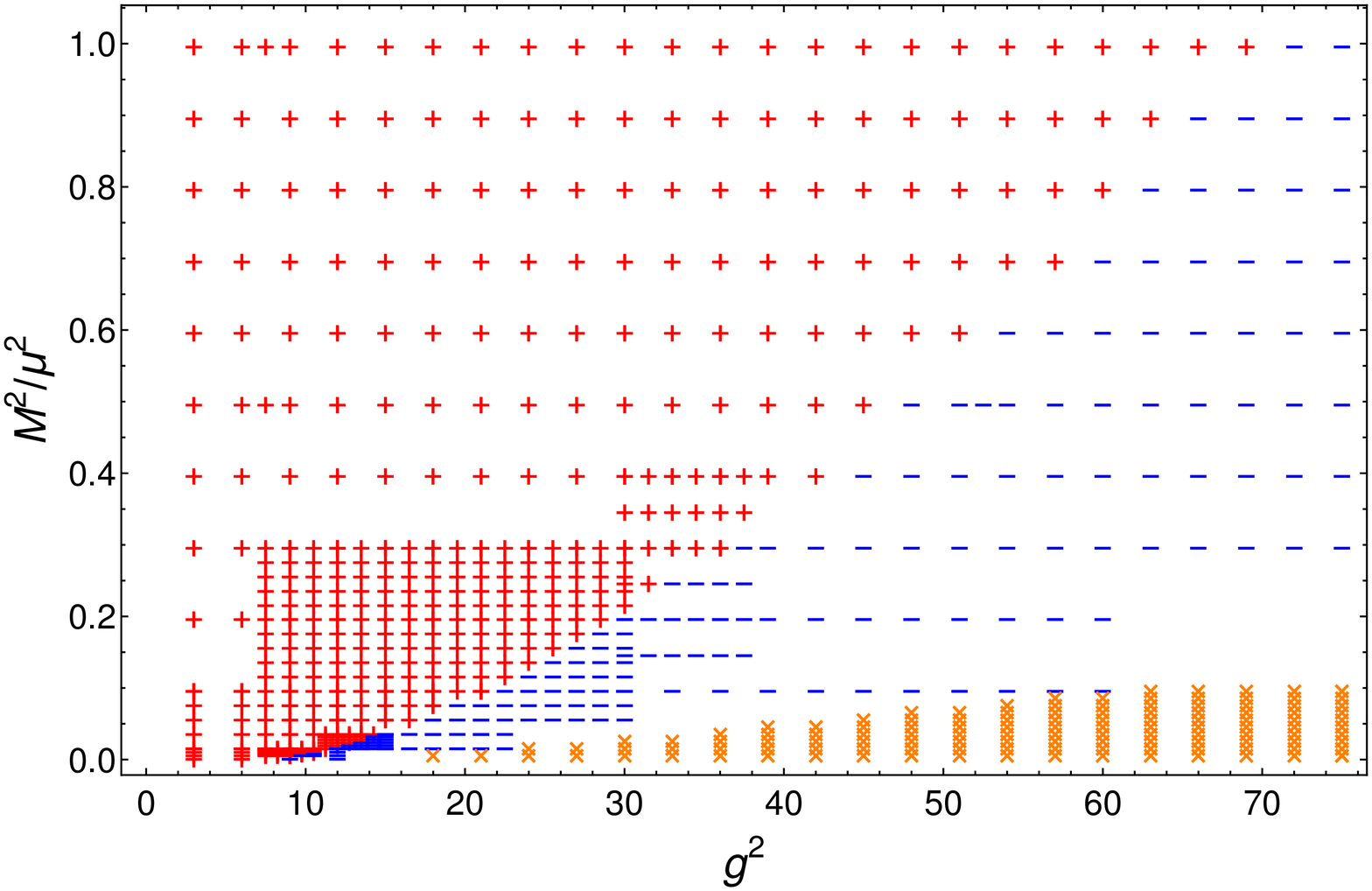}
\includegraphics[width=7.5cm]{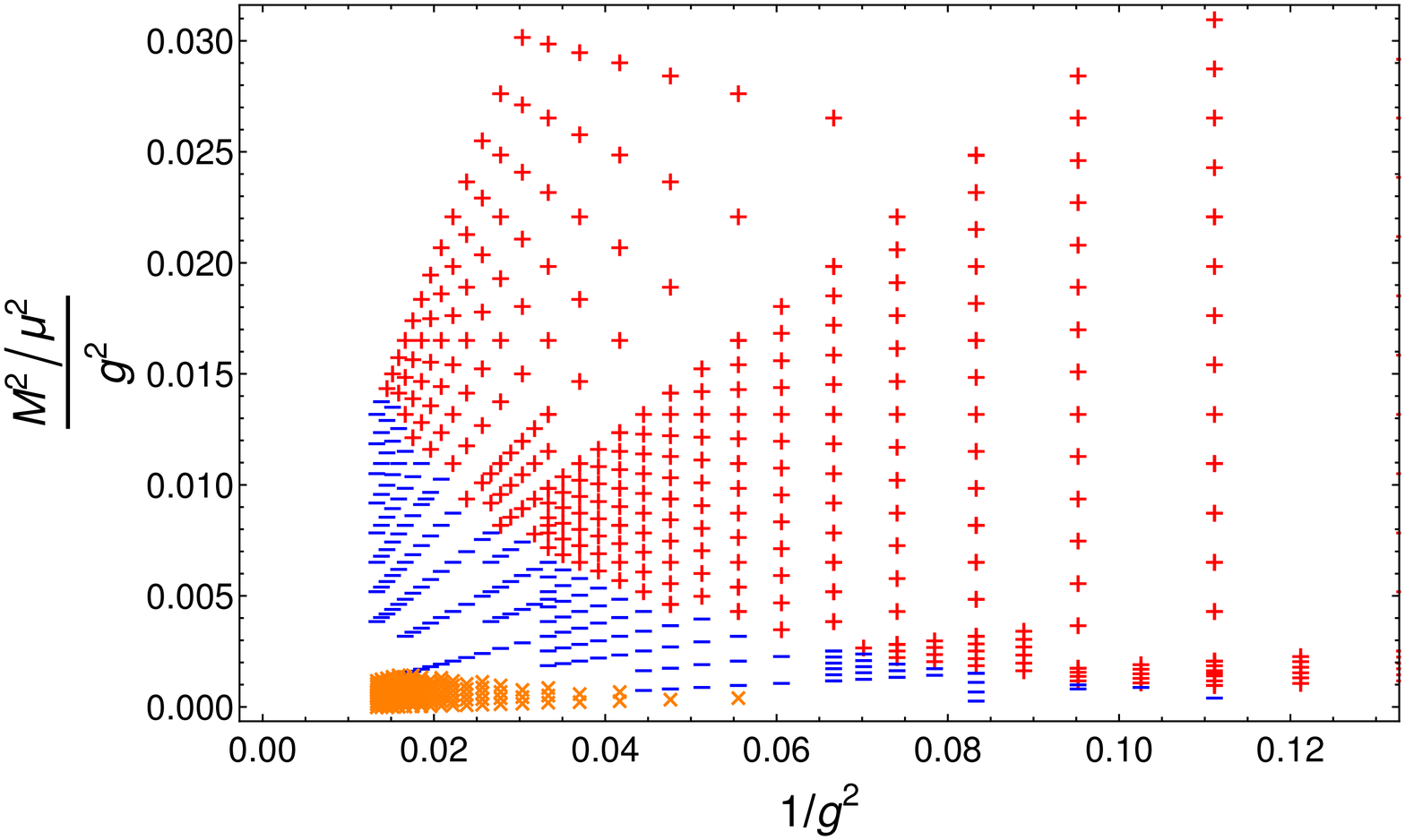}
%\vskip -0.7cm
\caption{\small
Distribution of positivity violation in the covariant Landau gauge $\alpha=0$ for the SU(2) Yang-Mills theory.% under the renormalization condition [TW]  
(left) in the plane $(g^2,\frac{M^2}{\mu^2})$ ,
(right) in the plane $(1/g^2,\frac{M^2/\mu^2}{g^2})$.
%The point in green indicates the physical value for the pure Yang-Mills theory. 
}
\label{cltw}
\end{figure}

We examine the magnitude of positivity violation in the two-dimensional phase diagram of the complementary $SU(2)$ gauge-scalar model.  
We draw a symbol $-$ in blue (positivity violation) on a point in the two-dimensional phase plane  $(g^2,\frac{M^2}{\mu^2})$ or $(1/g^2,\frac{M^2/\mu^2}{g^2})$ if the Schwinger function $\Delta(t)$  becomes negative in some regions of $t$ at the point, as indicated in Fig.~\ref{nonPos}.
We draw a symbol $+$ in red (positivity restoration) on a point in the two-dimensional phase plane  $(g^2,\frac{M^2}{\mu^2})$ 
or $(1/g^2,\frac{M^2/\mu^2}{g^2})$ if the Schwinger function $\Delta(t)$ is non-negative for any $t \in \mathbb{R}$ at the point, as indicated in Fig.~\ref{pos}. 
%The point in green indicates the physical value for the pure Yang-Mills theory. 
The symbol $\times$ in orange is used to show the point on which the solution is not reliable to be excluded from  considerations. %, as indicated in Fig.~\ref{singular}. 
Fig.~\ref{cltw} is the plot for the distribution of positivity violation in the covariant Landau gauge  $\alpha=0$ for the SU(2) Yang-Mills theory under the renormalization condition [TW].
%\noindent
%$\bullet$ 
If $g^2 \rightarrow 0$, the theory has no interaction and the propagator is nothing but the free massive propagator $D(k) = \frac{1}{k^2 + M^2}$.  
%Therefore, the Schwinger function must be positive in the small $g^2$ region for any value of $M$.
%In fact, our results are consistent with this observation.
% and there is no contribution of the loop calculations (e.g. $\Pi(k)$).
%\noindent
%$\bullet$ 
For small $g^2$ and large $M^2/\mu^2$, namely, for large $1/g^2$ and large $v^2 \simeq (M^2/\mu^2)/g^2$, the Schwinger function exhibits small violation of positivity. This region corresponds to the \textbf{Higgs-like region} in the gauge-scalar  model.
For large $g^2$ and small $M^2/\mu^2$, namely, for small $1/g^2$ and small $v^2 \simeq (M^2/\mu^2)/g^2$, the Schwinger function exhibits large violation of positivity.  This region corresponds to the \textbf{Confinement-like region} in the gauge-scalar  model.

%%%%%%%%%%%%%%%%%%%%%%%%%%%%%%%%%%%%%%%%%%%%%%%%%%%%%%%%%%%%%
%%%%%%%%%%%%%%%%%%%%%%%%%%%%%%%%%%%%%%%%%%%%%%%%%%%%%%%%%%%%%
\section{
Conclusion and discussion
}
%%%%%%%%%%%%%%%%%%%%%%%%%%%%%%%%%%%%%%%%%%%%%%%%%%%%%%%%%%%%%
%%%%%%%%%%%%%%%%%%%%%%%%%%%%%%%%%%%%%%%%%%%%%%%%%%%%%%%%%%%%%

\noindent
%$\bigodot$
We have examined the mass-deformed Yang-Mills theory in the covariant Landau gauge (which we called the massive Yang-Mills model) in order to reproduce the decoupling solution  as the  confining solution of the pure Yang-Mills theory. 
Indeed, the massive Yang-Mills model well reproduces the decoupling solution for  gluon and ghost propagators (at least) in the low-momentum region by choosing a suitable set of two parameters $g$ and $M$.

The massive Yang-Mills model in the covariant Landau gauge has the gauge-invariant extension, namely, the complementary gauge-scalar model with a radially fixed fundamental scalar field which is subject to an appropriate reduction condition. 
In other words, the gauge-scalar model with a radially fixed fundamental scalar field subject to the reduction condition can be gauge-fixed to become equal to the mass-deformed Yang-Mills theory in the covariant Landau gauge. 
The gauge-invariant extension of a non-gauge theory is performed through the gauge-independent description of the BEH mechanism  \cite{Kondo18} without relying on the spontaneous symmetry breaking which was first proposed for the adjoint scalar field \cite{Kondo16}. 
More details will be given in \cite{KSOMH18,HK18}.

\section*{Acknowledgement}

K.-I. K. was supported by Grant-in-Aid for Scientific Research, JSPS KAKENHI Grant Number (C) No.15K05042.
R. M. was supported by Grant-in-Aid for JSPS Research Fellow Grant Number 17J04780.

\end{document}